\documentclass{PoS}

\usepackage[utf8]{inputenc}
\usepackage{subcaption}
\usepackage{amsmath}
\usepackage{bm}
\usepackage{slashed}
\usepackage{verbatim}
\usepackage{multirow}
\usepackage{braket}
\usepackage{cite}

\newcommand{\st}{{\scriptscriptstyle T}}
\def\nn{\nonumber}
\def\cd{{\cdot}}
\DeclareMathOperator{\tr}{Tr}

\usepackage[normalem]{ulem} 
\renewcommand\sout{\bgroup \color[rgb]{0.55,0.00,0.99} \ULdepth=-.5ex \ULset}

\title{Gluon TMDs in the small-$\bm{x}$ limit}
\ShortTitle{Gluon TMDs in the small-$x$ limit}

\author{Tom van Daal$^{a,b}$ \\ \\
$^a$ Department of Physics and Astronomy, VU University Amsterdam, De Boelelaan 1081, NL-1081 HV Amsterdam, The Netherlands \\
$^b$ Nikhef, Science Park 105, NL-1098 XG Amsterdam, The Netherlands \\ \\
E-mail: \email{tvdaal@nikhef.nl}}
        
\abstract{In high-energy scattering processes involving two or more hadrons one can measure observables that are sensitive to partonic transverse momentum, which is encoded in so-called transverse momentum dependent (TMD) parton distribution functions (PDFs), also called TMDs. These functions correspond to Fourier transforms of matrix elements that contain process-dependent gauge links. As the energy associated to the collision process increases, one becomes more sensitive to the small-$x$ region which is dominated by gluon rather than quark TMDs. In this paper we study the leading-twist gluon TMDs in the small-$x$ limit for the dipole-type gauge link structure, for both unpolarized and vector polarized hadrons. In the limit $x\to0$, the gluon-gluon correlator reduces to a correlator that consists of a single Wilson loop. This is used to obtain a simple description of gluon TMDs in the small-$x$ region: some of the functions vanish, while others become proportional to each other.}

\FullConference{QCD Evolution 2016 \\
May 30 - June 3, 2016 \\
National Institute for Subatomic Physics (Nikhef) in Amsterdam}

\begin{document}

\section{Introduction}
To describe collisions involving hadrons in particle accelerators such as the Large Hadron Collider (LHC), knowledge is required of the internal structure of hadrons. The distribution of partons inside hadrons cannot be calculated perturbatively in quantum chromodynamics (QCD). Rather, this non-perturbative information needs to be extracted from data and is encoded in so-called parton distribution functions (PDFs). For inclusive processes such as deep-inelastic scattering (DIS), the cross section involves collinear PDFs that are functions of the longitudinal momentum fraction $x$ of the parton only. However, for less inclusive processes such as semi-inclusive DIS or Drell-Yan scattering, one can in addition become sensitive to partonic transverse momenta. The corresponding parton density functions are called transverse momentum dependent (TMD) PDFs (also simply called TMDs). Besides a dependence on $x$ and the transverse momentum $k_\st^2$, TMDs also depend on Wilson lines (also called gauge links). The gauge link is a necessary ingredient in any TMD correlator to achieve color gauge invariance. Gauge links are process dependent and thus give rise to a process dependence of the TMDs. The gluon-gluon TMD correlator, its gauge link structure, as well as its parametrization in terms of TMDs is discussed in section~\ref{s:correlator}.

Increasing the center-of-mass energy of hadronic collisions means probing smaller regions of $x$. For small $x$, gluon functions significantly dominate over the quark ones. In this paper we study the small-$x$ limit of gluon TMDs for spin-$1/2$ hadrons (spin-$0$ is automatically covered too by considering unpolarized hadrons). This analysis, as well as in part the one in ref.~\cite{Boer:2016dlh}, is based on a more detailed study presented in ref.~\cite{Boer:2016xqr}. In ref.~\cite{Boer:2016xqr} also tensor polarized targets (relevant for spin-$1$ hadrons) are discussed. We only consider TMDs at leading power in the inverse hard scale (also referred to as leading twist). Furthermore, we constrain ourselves to a particular choice of the gauge link, studying the so-called dipole-type TMDs. As it turns out, at $x=0$ the gluon-gluon correlator reduces to the Fourier transform of a hadronic matrix element containing a Wilson loop. This is the topic of section~\ref{s:limit}. The Wilson loop correlator, in turn, can also be parametrized in terms of TMDs. In the small-$x$ limit, the set of gluon TMDs simplifies to a much smaller number of Wilson loop TMDs. 

In this paper the hadron and parton momenta are denoted by $P$ and $k$ respectively. We parametrize $k$ in terms of the dimensionful vectors $P$ and $n$, where $n$ is a lightlike vector satisfying $n^2 = 0$ and $P \cd n = 1$:
\begin{equation}
    k^\mu = x P^\mu + k_\st^\mu + (k\cd P - xM^2) \,n^\mu , 
\end{equation}
where $M$ is the mass of the hadron and $P^2 = M^2$. Since we are interested in spin-$1/2$ hadrons, a spin vector $S$ is needed to describe vector polarized states. Also $S$ can be parametrized in terms of $P$ and $n$:
\begin{equation}
    S^\mu = S_L \frac{P^\mu}{M} + S_T^\mu - MS_L \,n^\mu ,
\end{equation}
satisfying $P \cd S = 0$.

\section{The gluon-gluon correlator} \label{s:correlator}
The gluon-gluon TMD (light-front) correlator for a spin-$1/2$ hadron is defined as
\begin{equation}
    \Gamma^{[U,U^\prime]\,\mu\nu;\rho\sigma}(x,\bm{k}_\st) \equiv \int \left. \frac{d\xi\cd P\, d^2\xi_\st}{(2\pi)^3} \;e^{ik\cdot\xi} \bra{P,S} \tr_c \left( F^{\mu\nu}(0) U_{[0,\xi]}^{\phantom{\prime}} F^{\rho\sigma}(\xi) U_{[\xi,0]}^\prime \right) \ket{P,S} \right|_{\xi\cd n=0} .
\end{equation}
This correlator contains the gauge links $U_{[0,\xi]}^{\phantom{\prime}}$ and $U_{[\xi,0]}^\prime$ in the fundamental representation of $\text{SU}(3)$, ensuring color gauge invariance. The path integrations of the gauge links are process dependent. The leading contributions to the correlator can be found by counting powers of $P \propto Q$ and $n \propto 1/Q$, where $Q$ denotes the hard scale. To leading power (also called leading twist), this correlator is given by
\begin{equation}
    \Gamma^{ij}(x,\bm{k}_\st) \equiv \Gamma^{[U,U^\prime]\,ni;nj}(x,\bm{k}_\st) ,
    \label{e:ltcorrelator}
\end{equation}
denoting contractions with the $n$-vector as superscripts and suppressing for convenience the gauge link dependence. Separating the various hadronic polarization states, this correlator can be parametrized in terms of TMDs as follows~\cite{Mulders:2000sh}:
\begin{equation}
    \Gamma^{ij}(x,\bm{k}_\st) = \Gamma_U^{ij}(x,\bm{k}_\st) + \Gamma_L^{ij}(x,\bm{k}_\st) + \Gamma_T^{ij}(x,\bm{k}_\st) ,
    \label{e:gamma_vp_tmds}
\end{equation}
where\footnote{In this paper momenta indicated in boldface are two-dimensional vectors on the transverse plane rather than four-vectors. We define $k_\st^\mu = [0,0,\bm{k}_\st]$ etc., so that e.g.\ $\bm{k}_\st \cd \bm{S}_\st = - k_\st \cd S_\st$.}
\begin{align}
    \Gamma_U^{ij}(x,\bm{k}_\st) &= \frac{x}{2} \left[ - \,g_\st^{ij} \,f_1(x,\bm{k}_\st^2) + \frac{k_\st^{ij}}{M^2} \,h_1^{\perp}(x,\bm{k}_\st^2) \right] , \\
    \Gamma_L^{ij}(x,\bm{k}_\st) &= \frac{x}{2} \left[ i \epsilon_\st^{ij} S_L \,g_1(x,\bm{k}_\st^2) + \frac{{\epsilon_\st^{\{i}}_\alpha k_\st^{j\}\alpha} S_L}{2M^2} \,h_{1L}^\perp(x,\bm{k}_\st^2) \right] , \\
    \Gamma_T^{ij}(x,\bm{k}_\st) &= \frac{x}{2} \left[ - \,\frac{g_\st^{ij} \epsilon_\st^{S_\st k_\st}}{M} \,f_{1T}^\perp(x,\bm{k}_\st^2) + \frac{i \epsilon_\st^{ij} \bm{k}_\st \cd \bm{S}_\st}{M} \,g_{1T}(x,\bm{k}_\st^2) \right. \nn \\
    &\quad\quad\;\; \left. - \,\frac{\epsilon_\st^{k_\st\{i} S_\st^{j\}} + \epsilon_\st^{S_\st\{i} k_\st^{j\}}}{4M} \,h_1(x,\bm{k}_\st^2) - \frac{{\epsilon_\st^{\{i}}_\alpha k_\st^{j\}\alpha S_\st}}{2M^3} \,h_{1T}^\perp(x,\bm{k}_\st^2) \right] ,
\end{align}
where $g_\st^{\mu\nu} \equiv g^{\mu\nu} - P^{\{\mu} n^{\nu\}}$ (curly brackets denote symmetrization of the indices), with nonvanishing elements $g_\st^{11} = g_\st^{22} = -1$, and $\epsilon_\st^{\mu\nu} \equiv \epsilon^{Pn\mu\nu}$, with nonzero components $\epsilon_\st^{12} = -\epsilon_\st^{21} = 1$. In the parametrization we have employed symmetric traceless tensors in $k_\st$, which ensures having TMDs of \emph{definite rank}. This is useful for studying them in impact parameter space, which is relevant for TMD evolution. The symmetric traceless tensors are defined in appendix~C of ref.~\cite{Boer:2016xqr}. The rank-$1$ function $h_1$ is related to the often used indefinite-rank function $h_{1T}$ as 
\begin{equation}
    h_1(x,\bm k_\st^2) \equiv h_{1T}(x,\bm k_\st^2) + \frac{\bm{k}_\st^2}{2M^2} \,h_{1T}^\perp(x,\bm k_\st^2) .
\end{equation}

\section{The small-$\bm{x}$ limit} \label{s:limit}
\begin{figure}[!htb]
    \centering
    \begin{subfigure}[b]{0.31\textwidth}
        \includegraphics[width=\textwidth]{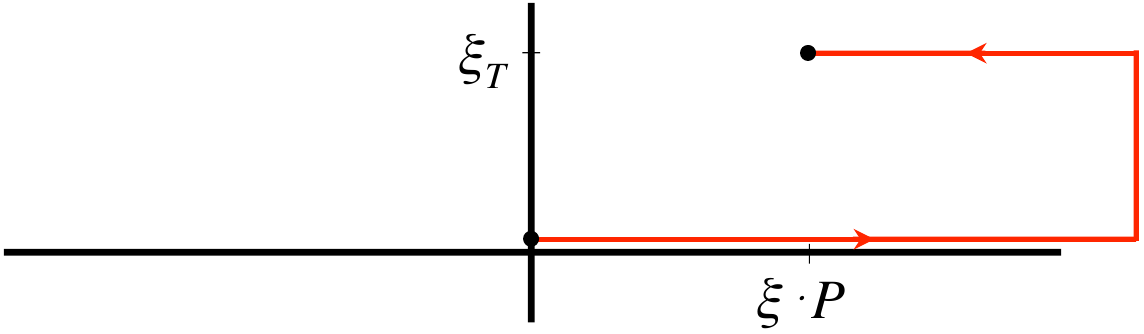} 
        \caption{}
        \label{f:plus}
    \end{subfigure} \quad
    \begin{subfigure}[b]{0.31\textwidth}
        \includegraphics[width=\textwidth]{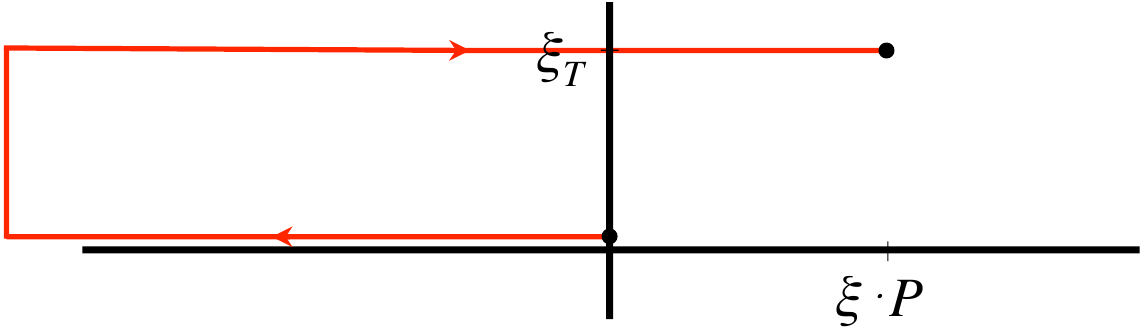} 
        \caption{}
        \label{f:minus}
    \end{subfigure} \quad
    \begin{subfigure}[b]{0.31\textwidth}
        \includegraphics[width=\textwidth]{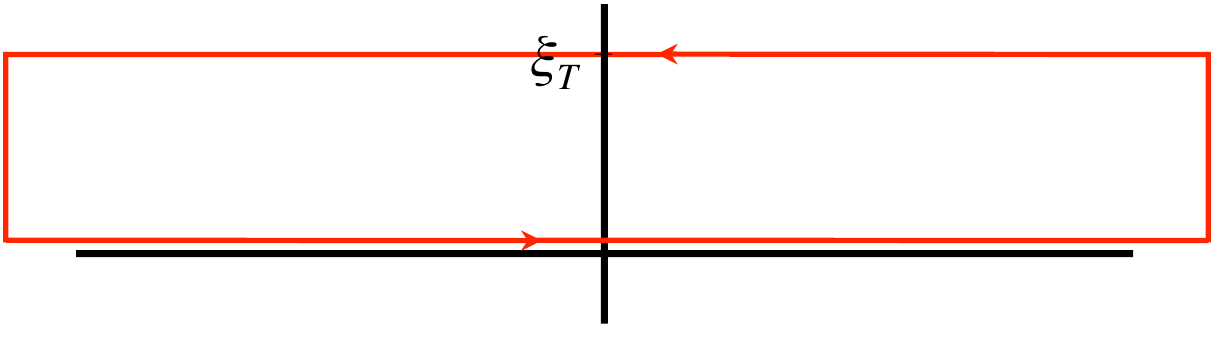} 
        \caption{}
        \label{f:loop}
    \end{subfigure}      
    \caption{The so-called plus and minus staple-like gauge links (a) $U_{[0,\xi]}^{[+]}$ and (b) $U_{[0,\xi]}^{[-]}$ run along the light-front ($\xi\cd n = 0$) via plus or minus light cone infinity respectively, connecting the points $0$ and $\xi$. The plus and minus gauge links make up the Wilson loop (c) $U^{[\Box]} \equiv U_{[0,\xi]}^{[+]} U_{[\xi,0]}^{[-]} = U_{[0,\xi]}^{[+]} U_{[0,\xi]}^{[-]\dag}$, with transverse extent $\xi_\st$ (the longitudinal extent is infinite).}
    \label{f:links}
\end{figure}

Let us now consider for the gluon-gluon correlator the so-called dipole-type gauge link structure, which consists of a future-pointing $U_{[0,\xi]}^{[+]}$ and a past-pointing $U_{[\xi,0]}^{[-]}$ staple-like gauge link (see fig.~\ref{f:links}). 
To study the dipole-type gluon-gluon correlator at small $x$, we use the result obtained in ref.~\cite{Boer:2016xqr}:
\begin{align}
    \Gamma^{[+,-]\,ij}(0,\bm{k}_\st) &= \int \left. \frac{d\xi\cd P\,d^2\xi_\st}{(2\pi)^3} \;e^{ik\cdot\xi} \bra{P,S} \tr_c \left( F^{ni}(0)\,U^{[+]}_{[0,\xi]}\,F^{nj}(\xi)\,U^{[-]}_{[\xi,0]} \right) \ket{P,S}\right|_{\xi\cd n = k\cd n = 0} \nn \\
    &= \frac{k_\st^i k_\st^j}{2\pi L} \;\Gamma_0^{[\Box]}(\bm{k}_\st) ,
    \label{e:correlatorrel}
\end{align}
where the so-called Wilson loop correlator
\begin{equation}
    \Gamma_0^{[\Box]}(\bm{k}_\st) \equiv \left. \int \frac{d^2\xi_\st}{(2\pi)^2} \;e^{ik_\st\cdot\xi_\st} \bra{P,S} \tr_c \left( U^{[\Box]} \right) \ket{P,S}\right|_{\xi\cd n = 0} , 
    \label{e:WLcorrelator}
\end{equation}
contains a rectangular Wilson loop $U^{[\Box]} \equiv U_{[0,\xi]}^{[+]} U_{[\xi,0]}^{[-]}$ (see fig.~\ref{f:links}). Note in eq.~\eqref{e:correlatorrel} the proportionality to the longitudinal dimension $L$ of the Wilson loop, $L \equiv \int d\xi\cd P = 2\pi\,\delta(0)$. Assuming continuity in $x$, eq.~\eqref{e:correlatorrel} implies that
\begin{equation}
    \lim_{x\to0} \;\Gamma^{[+,-]\,ij}(x,\bm{k}_\st) = \frac{k_\st^i k_\st^j}{2\pi L} \;\Gamma_0^{[\Box]}(\bm{k}_\st) .
    \label{e:smallxrel}
\end{equation}

Also the Wilson loop correlator in eq.~\eqref{e:WLcorrelator} can be parametrized in terms of TMDs. We denote them by the letter `$e$' to distinguish them from the familiar $f,g,h$-type TMDs. Constrained by hermiticity and parity conservation, a possible parametrization is given by~\cite{Boer:2016xqr}
\begin{equation}
    \Gamma_0^{[\Box]}(\bm{k}_\st) = \Gamma_{0\hspace{0.08cm}U}^{[\Box]}(\bm{k}_\st) + \Gamma_{0\hspace{0.08cm}L}^{[\Box]}(\bm{k}_\st) + \Gamma_{0\hspace{0.08cm}T}^{[\Box]}(\bm{k}_\st) ,
    \label{e:gamma0_vp_tmds}
\end{equation}
where
\begin{align}
    \Gamma_{0\hspace{0.08cm}U}^{[\Box]}(\bm{k}_\st) &= \frac{\pi L}{M^2} \,e(\bm{k}_\st^2) , \\
    \Gamma_{0\hspace{0.08cm}L}^{[\Box]}(\bm{k}_\st) &= 0 , \\
    \Gamma_{0\hspace{0.08cm}T}^{[\Box]}(\bm{k}_\st) &= \frac{\pi L}{M^2} \,\frac{\epsilon_\st^{S_\st k_\st}}{M} \,e_T(\bm{k}_\st^2) .
\end{align}

In the small-$x$ limit eq.~\eqref{e:smallxrel} relates the gluon TMDs in eq.~\eqref{e:gamma_vp_tmds} to the Wilson loop TMDs in eq.~\eqref{e:gamma0_vp_tmds}. For the unpolarized case this limit gives
\begin{equation}
    \lim_{x\to0} \;\Gamma_U^{ij}(x,\bm{k}_\st) = \frac{1}{2} \left( - \,g_\st^{ij} \,\frac{\bm k_\st^2}{2M^2} + \frac{k_\st^{ij}}{M^2} \right) e(\bm{k}_\st^2) ,
\end{equation}
implying that
\begin{equation}
    \lim_{x\to0} \,x f_1(x,\bm{k}_\st^2) = \frac{\bm{k}_\st^2}{2M^2} \lim_{x\to0} \,x h_1^\perp(x,\bm{k}_\st^2) = \frac{\bm{k}_\st^2}{2M^2} \,e(\bm{k}_\st^2) .
    \label{e:upol_result}
\end{equation} 
Note that $h_1^\perp$ saturates its positivity bound~\cite{Mulders:2000sh}. For longitudinally polarized hadrons eq.~\eqref{e:smallxrel} implies that $g_1$ and $h_{1L}^\perp$ are suppressed in the small-$x$ limit. For transversely polarized hadrons we find
\begin{equation}
    \lim_{x\to0} \;\Gamma_T^{ij}(x,\bm{k}_\st) = \frac{1}{2} \left( - \,\frac{g_\st^{ij} \epsilon_\st^{S_\st k_\st}}{M} \,\frac{\bm{k}_\st^2}{2M^2} - \frac{\epsilon_\st^{k_\st\{i} S_\st^{j\}} + \epsilon_\st^{S_\st\{i} k_\st^{j\}}}{4M} \,\frac{\bm{k}_\st^2}{2M^2} + \frac{{\epsilon_\st^{\{i}}_\alpha k_\st^{j\}\alpha S_\st}}{2M^3} \right) e_T(\bm{k}_\st^2) ,
\end{equation}
implying that
\begin{align}
    \lim_{x\to0} \,x f_{1T}^\perp(x,\bm{k}_\st^2) &= \lim_{x\to0} \,x h_1(x,\bm{k}_\st^2) = - \frac{\bm{k}_\st^2}{2M^2} \lim_{x\to0} \,x h_{1T}^\perp(x,\bm{k}_\st^2) = \frac{\bm{k}_\st^2}{2M^2} \,e_T(\bm{k}_\st^2) ,
    \label{e:tpol_result}
\end{align}
while $x g_{1\st}(x,\bm{k}_\st^2)$ vanishes in this limit\footnote{Since the right-hand side of eq.~\eqref{e:smallxrel} is symmetric in $i,j$, so must the left-hand side. The $g$-type functions, by definition, are accompanied by the Lorentz structure $\epsilon_\st^{ij}$, which is antisymmetric in $i,j$. Hence, regardless of the hadronic polarization state, all $g$-type functions vanish in this small-$x$ limit analysis.}. This is in agreement with the results in ref.~\cite{Boer:2015pni} where the small-$x$ limit was studied for transversely polarized hadrons. The function $e_T$ is related to what in literature is known as the (spin-dependent) odderon operator~\cite{Boer:2015pni,Hatta:2005as,Zhou:2013gsa,Szymanowski:2016mbq}.

In table~\ref{t:summary} we list the leading-twist TMDs (multiplied by $x$) and the $x\rightarrow 0$ limit in the case of the dipole-type gauge link structure (i.e. one future- and one past-pointing Wilson line). Some TMDs are expected to be zero at $x = 0$, while others become proportional to (moments of) the Wilson loop $e$-type TMDs. The nonvanishing functions grow as $1/x$ towards small $x$ up to subdominant corrections coming from resummed logarithms of $1/x$. Furthermore, the behavior of the TMDs under time reversal ($T$) and charge conjugation ($C$) is listed\footnote{See appendix~A in ref.~\cite{Boer:2016xqr} for the precise definitions of $T$ and $C$.}. The $T$- and $C$-behavior of the functions nicely matches in the small-$x$ limit. \\
\begin{table}[!htb]
    \centering
    {\renewcommand{\arraystretch}{1.2}
    \begin{tabular}{|c|c|c|c|c|c|}
    \hline
    & \textbf{Rank} & $\bm{T}$ & $\bm{C}$ & \textbf{Limit} $\bm{x \to 0}$ \\ \hline \hline
    $xf_1$ & $0$ & even & even & $e^{(1)}$ \\
    $xh_1^\perp$ & $2$ & even & even & $e$ \\ \hline
    $xg_1$ & $0$ & even & odd & $0$ \\
    $xh_{1L}^\perp$ & $2$ & odd & even & $0$ \\
    $xf_{1T}^\perp$ & $1$ & odd & odd & $e_T^{(1)}$ \\
    $xg_{1T}$ & $1$ & even & even & $0$ \\
    $xh_{1}$ & $1$ & odd & odd & $e_T^{(1)}$ \\
    $xh_{1T}^\perp$ & $3$ & odd & odd & $-e_T$ \\ \hline
    \end{tabular}}
    \caption{An overview of the leading-twist gluon TMDs for unpolarized and vector polarized hadrons. In the second column we list the rank of the function. The rank-$0$ functions also appear as collinear PDFs. In the next columns we list the properties (even/odd) under time reversal ($T$) and charge conjugation ($C$). In the last column it is indicated to which $e$-type function the TMD reduces in the limit $x \to 0$. For convenience we use the moment notation $f_{\ldots}^{(n)}(x,\bm{k}_\st^2) \equiv [\bm{k}_\st^2/(2M^2)]^n \,f_{\ldots}(x,\bm{k}_\st^2)$.}
    \label{t:summary}
\end{table}

\section{Conclusion}
We have studied the gluon-gluon TMD correlator for unpolarized and vector polarized hadrons. To leading power in the inverse hard scale, this correlator can be parametrized in terms of leading-twist TMDs. Choosing the dipole-type gauge link structure, we have shown that in the small-$x$ limit the gluon-gluon correlator reduces to the Wilson loop correlator, which in turn can also be parametrized in terms of TMDs. The gluon TMDs either vanish when $x\to0$ or become proportional to the Wilson loop TMDs and then scale as $1/x$ (up to resummed logarithms), leading to a very simple picture of gluon functions in the small-$x$ region.

\acknowledgments
The presented work has been performed in collaboration with Dani\"el Boer, Sabrina Cotogno, Piet J. Mulders, Andrea Signori, and Ya-Jin Zhou. This research is part of the research program of the ``Stichting voor Fundamenteel Onderzoek der Materie (FOM)'', which is financially supported by the ``Nederlandse Organisatie voor Wetenschappelijk Onderzoek (NWO)'' as well as the EU FP7 ``Ideas'' programme QWORK (contract no. 320389).

\bibliography{references}

\providecommand{\href}[2]{#2}\begingroup\raggedright\begin{thebibliography}{1}

\bibitem{Boer:2016dlh}
D.~Boer, S.~Cotogno, T.~van Daal, P.~J. Mulders, A.~Signori and Y.-J. Zhou,
  \emph{{Gluon transverse momentum dependent correlators in polarized high
  energy processes}},  in \emph{{Proceedings, 24th International Workshop on
  Deep-Inelastic Scattering and Related Subjects (DIS 2016): Hamburg, Germany,
  April 11-25, 2016}}, 2016.
\newblock \href{http://arxiv.org/abs/1609.02788}{{\tt 1609.02788}}.

\bibitem{Boer:2016xqr}
D.~Boer, S.~Cotogno, T.~van Daal, P.~J. Mulders, A.~Signori and Y.-J. Zhou,
  \emph{{Gluon and Wilson loop TMDs for hadrons of spin $\leq$ 1}},
  \href{http://arxiv.org/abs/1607.01654}{{\tt 1607.01654}}.

\bibitem{Mulders:2000sh}
P.~J. Mulders and J.~Rodrigues, \emph{Transverse momentum dependence in gluon
  distribution and fragmentation functions}, {\emph{Phys. Rev.} {\bf D63}
  (2001) 094021}, [\href{http://arxiv.org/abs/hep-ph/0009343}{{\tt
  hep-ph/0009343}}].

\bibitem{Boer:2015pni}
D.~Boer, M.~G. Echevarria, P.~Mulders and J.~Zhou, \emph{{Single spin
  asymmetries from a single Wilson loop}},
  \href{http://dx.doi.org/10.1103/PhysRevLett.116.122001}{\emph{Phys. Rev.
  Lett.} {\bf 116} (2016) 122001}, [\href{http://arxiv.org/abs/1511.03485}{{\tt
  1511.03485}}].

\bibitem{Hatta:2005as}
Y.~Hatta, E.~Iancu, K.~Itakura and L.~McLerran, \emph{{Odderon in the color
  glass condensate}},
  \href{http://dx.doi.org/10.1016/j.nuclphysa.2005.05.163}{\emph{Nucl. Phys.}
  {\bf A760} (2005) 172--207}, [\href{http://arxiv.org/abs/hep-ph/0501171}{{\tt
  hep-ph/0501171}}].

\bibitem{Zhou:2013gsa}
J.~Zhou, \emph{{Transverse single spin asymmetries at small x and the anomalous
  magnetic moment}},
  \href{http://dx.doi.org/10.1103/PhysRevD.89.074050}{\emph{Phys. Rev.} {\bf
  D89} (2014) 074050}, [\href{http://arxiv.org/abs/1308.5912}{{\tt
  1308.5912}}].

\bibitem{Szymanowski:2016mbq}
L.~Szymanowski and J.~Zhou, \emph{{The spin dependent odderon in the diquark
  model}},  \href{http://arxiv.org/abs/1604.03207}{{\tt 1604.03207}}.

\end{thebibliography}\endgroup
\bibliographystyle{JHEP}

\end{document}